\documentclass[reqno]{article}
\usepackage{alltt}
\usepackage{graphicx}
\begin{document}
\title{Field Equations of the CP(N-1) Affine Gauge Theory}
\author{P.Leifer$^{1a}$ and L.P.Horwitz$^{abc}$}
\date{$^a$Bar-Ilan University, Ramat-Gan, Israel; \\$^b$School of
Physics
and Astronomy, Tel-Aviv University, Ramat Aviv, Israel; \\
$^c$College of Judea and Shomron, Ariel, Israel} \maketitle
\footnotetext[1]{On leave from Crimea State Engineering and
Pedagogical University, Simferopol, Crimea, Ukraine}
\begin{abstract}
A non-linear relativistic 4D field model of a quantum particle which
emerges from the internal dynamics in the quantum phase space
$CP(N-1)$ is proposed. In this model there is no distinction between
`particle' and its `surrounding field', and the space-time manifold
emerges from the description of the quantum state. The quantum
observables of the `quantum particle field' are described in terms
of the affine parallel transport of the local dynamical variables in
$CP(N-1)$.
\end{abstract}
\vskip 0.1cm
\section{Introduction}
We discuss in this paper a simple model of affine gauge field
equations describing some generalized coherent state. This model
admits both Goldstone and Higgs modes of evolution, whose dynamical
properties may be described in terms of nonlinear relativistic wave
equations. The model, furthermore, has the property that the
physical particle includes, by definition, its self-interaction in a
natural way. The basic structure of the model, involving the natural
motions in a $CP(N-1)$ manifold induced by the generators of $SU(N)$
has been discussed elsewhere \cite{Le1,Le2,Le3}. The invariant
Cartan decomposition of the Lie algebra $Alg SU(N)$ of generators
expressed in terms of the local coordinates admits an invariant
embedding of the isotropy sub-group $H=U(1)\times U(N-1) \subset
G=SU(N)$ and the coset transformations $G/H=SU(N)/[U(1) \times
U(N-1)]=CP(N-1)$ of the generalized coherent state (GCS). The
symplectic and Riemannian structure of the $CP(N-1)$ manifold may be
expressed in local coordinates as well
\cite{CMP,Hughston1,Hughston2,AdHr}. The normal and tangent vectors
at a point on the $CP(N-1)$ manifold form the unitary basis of a
spinor corresponding to an observable associated with a two state
system, as we shall explain below. The affine connection associated
with the Fubini-Study metric gives an invariant relation between
such vectors at nearby points which can be described in terms of the
transformations of $SL(2,C)$. There is a close analogy between those
and Lorentz transformations in the actual space-time. In this
framework, we construct nonlinear relativistic field equations which
are covariant under the diffeomorphisms generated by the local
actions of $SU(N)$, realized in terms of the $SL(2,C)$ action on the
local spinor in the state space $C^2$. We review the basic structure
briefly  in the following, and show how the field equations arise.
We finally discuss some properties of the solutions of these
equations which could be associated with elementary particles.

The main new points of our approach are as follows:

A. We use the notion of ``elementary quantum motion states" with
well defined quantized Planck's action $S_a=\hbar a$. Their GCS
serve as an abstract formalization of the ``quasi-classical''
description of a quantum setup or ``Schr\"odinger's lump"
\cite{Penrose}.

B. The quantum phase space $CP(N-1)$ serves as the base of a fibre
bundle for which the tangent space corresponds to local dynamical
variables (LDV's). The particular section of this bundle and the
corresponding affine gauge field are geometric tools for the
construction of a theory admitting quantum measurement in the
state-dependent dynamical space-time.

The technical details are as follows:

1. The projective representation of pure $N$-dimension quantum
states (one could think of arbitrary large $N$), provides a natural
non-linear realization of the $G=SU(N)$ group manifold and the coset
sub-manifold $G/H = SU(N)/S[U(1) \times U(N-1)]=CP(N-1)$. We shall
consider the generators of this group as local dynamical variables
\cite{Le3} of the model.

2. These quantum dynamical variables are represented by the tangent
vector fields to $CP(N-1)$. Embedding of $CP(N-1)$ into
$\mathcal{H}=C^N$ provides a measurement procedure for the dynamical
variables.

3. This notion of measurement gives rise to the analog of a local
dynamical space-time capable of detecting the coincidence of
expectation and measured values of these quantum dynamical
variables.

4. The affine parallel transport, associated with the Fubini-Study
metric, accompanied with ``Lorentz spin transformation matrix"
\cite{G}, establish this coincidence due to the identification of
the parallel transported LDV in different GCS's.

5. The parametrization of the measurement, with the help of {\it
attributed} local space-time coordinates, is in fact an embedding of
quantum dynamics in Hilbert space into a 4D world. This procedure is
well defined due to the existence of the infinitesimal $SL(2,C)$
transformations of the spinor treated as Lorentz transformations of
local space-time coordinates.

Below are introduced some fundamental notions of our construction.

\section{The Action Quantization}
We discuss a modification of the ``second quantization'' procedure.

{\it First.} In the second quantization method one has formally
given particles whose properties are defined by some commutation
relations between creation-annihilation operators. Note, that the
commutation relations are only the simplest consequence of the
curvature of the dynamical group manifold in the vicinity of the
group's unit (in algebra). Dynamical processes require, however,
finite group transformations and, hence, the global group structure.
The main technical idea is to use vector fields over a group
manifold instead of Dirac's q-numbers. This scheme therefore seeks
the dynamical nature of the creation and annihilation processes of
quantum particles.

{\it Second.}  We shall primarily  quantize the action, and not the
energy. The relative (local) vacuum of some problem is not
necessarily the state with minimal energy, it is a state with an
extremal of some action functional.

POSTULATE 1.

\noindent {\it We assume that there are elementary quantum states
$|\hbar a>, a=0,1,...$ of an abstract Planck oscillator whose states
correspond to the quantum motions with given number of Planck action
quanta}.

We shall construct non-linear field equations describing energy
(frequency) distribution, whose soliton-like solution provides the
quantization of the dynamical variables. Quantum ``particles'', and,
hence, their numbers arise as some countable solutions of non-linear
wave equations. In order to establish acceptable field equations
which are capable of intrinsically describing all possible degrees
of freedom under intensive interaction we construct a {\it universal
ambient Hilbert state space} $\cal{H}$. We will use {\it the
universality of the action} whose variation is capable of generating
any relevant dynamical variable. We shall call vectors of the action
state space $\cal{H}$ {\it action amplitudes}. Some of them will be
elementary quantum states of motion corresponding to discrete
numbers of Planck's quanta $| \hbar a>$. The action may create a
linear superposition of $|\hbar a>=(a!)^{-1/2} ({\hat \eta^+})^a|0>$
constituting $SU(\infty)$ multiplete of the Planck's action quanta
operator $\hat{S}=\hbar {\hat \eta^+} {\hat \eta}$ with the spectrum
$S_a=\hbar a$ in the separable Hilbert space $\cal{H}$. The standard
basis $\{|\hbar a>\}_0^{\infty}$ will be used with the `principle'
quantum number $a=0,1,2...$ assigned by Planck's quanta counting.
Generally action amplitudes are a coherent superposition
\begin{eqnarray}
|G>=\sum_{a=0}^{\infty} g^a| \hbar a>,
\end{eqnarray}
which may represent of the ground state, or ``vacuum", of some
quantum system. In fact only finite, say, $N$ elementary quantum
states may be involved. Then one may restrict $CP(\infty)$ to finite
dimensional $CP(N-1)$. Hereafter we will use the indices as follows:
$0\leq a \leq N$, and $1\leq i,k,m,n,s \leq N-1$

Since any ray of action amplitude has isotropy group $H=U(1)\times
U(N)$, in $\cal{H}$ only coset transformations $G/H=SU(N)/S[U(1)
\times U(N-1)]=CP(N-1)$ effectively act. Therefore the ray
representation of $SU(N)$ in $C^N$, in particular, the embedding of
$H$ and $G/H$ in $G$, is a state-dependent parametrization. Hence,
there is a diffeomorphism between the space of the rays marked by
the local coordinates in the map
 $U_j:\{|G>,|g^j| \neq 0 \}, j>0$
\begin{equation}
\pi^i_{(j)}=\cases{\frac{g^i}{g^j},&if $ 1 \leq i < j$ \cr
\frac{g^{i+1}}{g^j}&if $j \leq i < N-1$}
\end{equation}\label{coor}
and the group manifold of the coset transformations
$G/H=SU(N)/S[U(1) \times U(N-1)]=CP(N-1)$. This diffeomorphism is
provided by the coefficient functions $\Phi^i_{\alpha}$ of the local
generators (see below). The choice of the map $U_j$ means, that the
comparison of quantum amplitudes refers to the amplitude with the
action $\hbar j$. The breakdown of $SU(N)$ symmetry on each action
amplitude to the isotropy group $H=U(1)\times U(N-1)$ contracts the
full dynamics down to $CP(N-1)$. The physical interpretation of
these transformations is given by the

POSTULATE 2.

\noindent {\it We shall assume that the unitary transformations of
the action amplitudes may be identified with physical fields; i.e.,
transformations of the form}
$U(\tau)=\exp(i\Omega^{\alpha}\hat{\lambda}_{\alpha}\tau)$, where
the field functions $\Omega^{\alpha}$ are in the adjoint
representations of $SU(N)$ {\it The coset transformation
$G/H=SU(N)/S[U(1)\times U(N-1)]=CP(N-1)$ is the quantum analog of a
classical force; its action is equivalent to some physically
distinguishable variation of GCS in $CP(N-1)$}.

The $CP(N-1)$ manifold takes the place of the ``classical phase
space'' \cite{ZF}, since its points, corresponding to the GCS, are
most close to classical states of motion. Two interpretations may be
given for the points of $CP(N-1)$. One of them is the
``Schr\"odinger's lump" \cite{Penrose} and the second one is the
analog of the Stern-Gerlach ``filter's orientations" discussed by
Fivel \cite{Fivel}. The basic content of their physical
interpretations is that one has {\it a macroscopic (i.e. space-time)
discriminator} of two quantum states. As such, they may be used as
``yes/no'' states of some two-level detector. We will use the
``Schr\"odinger's lump" interpretation. Let us assume that GCS
described by local coordinates $(\pi^1,...,\pi^{N-1})$ corresponds
to the original lump, and the coordinates $(\pi^1+\delta
\pi^1,...,\pi^{N-1}+\delta \pi^{N-1})$ correspond to the lump
displaced due to measurement. Such coordinates of the lump gives the
a firm geometric tool for the description of quantum dynamics during
interaction which may used for a measuring process.

Then the question that we now want to raise is the following: {\it
what ``classical field'', i.e. field in space-time, corresponds to
the transition from the original to the displaced lump?} In other
words we would like to find the measurable physical manifestation of
the lump , which we shall call the ``field shell", its space-time
shape and its dynamics. The lump's perturbations will be represented
by ``geometric bosons'' \cite{Le4} whose frequencies are not a
priori given, but are defined by some field equations which should
established by means of a new variation problem. Before its
formulation, we wish to introduce differential geometric
construction.

We will assume that all ``vacua'' solutions belong to a single
separable {\it projective Hilbert space} $CP(N-1)$. The vacuum is
now the stationary point of some action functional, not a solution
with the minimal energy. Energy will be associated with vector field
tangent to $CP(N-1)$ giving the rate of change of the action
variation with respect to the notion of
Newton-Stueckelberg-Horwitz-Piron time (world time) \cite{H1}.
Dynamical space-time will be built at any GCS and, particularly, at
the vacuum of some ``classical'' problem (see below). Therefore
Minkowsky space-time appears to be functionally local
(state-dependent) in $CP(N-1)$ and the space-time motion dictated by
the field equations connected with two infinitesimally close
``vacua''. The connection between these local space-times may be
physically established by the measurement given in terms of geometry
of the base manifold $CP(N-1)$.

Now we are evidences of the so-called Multiverse (omnium) concept
\cite{W1,Penrose}. We think there is only one Universe but there
exists a continuum of dynamical space-times each of them related to
one point of the quantum phase space $CP(N-1)$. The standard
approach, identifying the Universe with a single space-time, appears
to be too strong an assumption from this point of view.

\section{Local dynamical variables}
The state space ${\cal H}$ of the field configurations with finite
action quanta is a stationary construction. We introduce dynamics
{\it by the velocities of the GCS variation} representing some
``elementary excitations'' (quantum particles). Their dynamics is
specified by the Hamiltonian, giving time variation velocities of
the action quantum numbers in different directions of the tangent
Hilbert space $T_{(\pi^1,...,\pi^{N-1})} CP(N-1)$ which takes the
place of the ordinary linear quantum scheme. The rate of the action
variation gives the energy of the ``particles''.

The local dynamical variables correspond to internal symmetries of
the GCS and their evolution should be expressed now in terms of the
local coordinates $\pi^k$. The Fubini-Study metric
\begin{equation}
G_{ik^*} = [(1+ \sum |\pi^s|^2) \delta_{ik}- \pi^{i^*} \pi^k](1+
\sum |\pi^s|^2)^{-2} \label{FS}
\end{equation}
and the affine connection
\begin{eqnarray}
\Gamma^i_{mn} = \frac{1}{2}G^{ip^*} (\frac{\partial
G_{mp^*}}{\partial \pi^n} + \frac{\partial G_{p^*n}}{\partial
\pi^m}) = -  \frac{\delta^i_m \pi^{n^*} + \delta^i_n \pi^{m^*}}{1+
\sum |\pi^s|^2} \label{Gamma}
\end{eqnarray}
in these coordinates will be used. Hence the internal dynamical
variables and their norms should be state-dependent, i.e. local in
the state space \cite{Le1,Le2,Le3,Le4}. These local dynamical
variables realize a non-linear representation of the unitary global
$SU(N)$ group in the Hilbert state space $C^N$. Namely, $N^2-1$
generators of $G = SU(N)$ may be divided in accordance with the
Cartan decomposition: $[B,B] \in H, [B,H] \in B, [B,B] \in H$. The
$(N-1)^2$ generators
\begin{eqnarray}
\Phi_h^i \frac{\partial}{\partial \pi^i}+c.c. \in H,\quad 1 \le h
\le (N-1)^2
\end{eqnarray}
of the isotropy group $H = U(1)\times U(N-1)$ of the ray (Cartan
sub-algebra) and $2(N-1)$ generators
\begin{eqnarray}
\Phi_b^i \frac{\partial}{\partial \pi^i} + c.c. \in B, \quad 1 \le b
\le 2(N-1)
\end{eqnarray}
are the coset $G/H = SU(N)/S[U(1) \times U(N-1)]$ generators
realizing the breakdown of the $G = SU(N)$ symmetry of the GCS.
Furthermore, the $(N-1)^2$ generators of the Cartan sub-algebra may
be divided into the two sets of operators: $1 \le c \le N-1$ ($N-1$
is the rank of $Alg SU(N)$) Abelian operators, and $1 \le q \le
(N-1)(N-2)$ non-Abelian operators corresponding to the
non-commutative part of the Cartan sub-algebra of the isotropy
(gauge) group. Here $\Phi^i_{\sigma}, \quad 1 \le \sigma \le N^2-1 $
are the coefficient functions of the generators of the non-linear
$SU(N)$ realization. They give the infinitesimal shift of the
$i$-component of the coherent state driven by the $\sigma$-component
of the unitary multipole field $\Omega^{\alpha}$ rotating the
generators of $Alg SU(N)$ and they are defined as follows:
\begin{equation}
\Phi_{\sigma}^i = \lim_{\epsilon \to 0} \epsilon^{-1}
\biggl\{\frac{[\exp(i\epsilon \lambda_{\sigma})]_m^i g^m}{[\exp(i
\epsilon \lambda_{\sigma})]_m^j g^m }-\frac{g^i}{g^j} \biggr\}=
\lim_{\epsilon \to 0} \epsilon^{-1} \{ \pi^i(\epsilon
\lambda_{\sigma}) -\pi^i \},
\end{equation}
\cite{Le5}. Then the sum of the $N^2-1$ the energies associated with
intensity of deformations of the GCS is represented  by the local
Hamiltonian vector field $H$ which is linear in the partial
derivatives $\frac{\partial }{\partial \pi^i} = \frac{1}{2}
(\frac{\partial }{\partial \Re{\pi^i}} - i \frac{\partial }{\partial
\Im{\pi^i}})$ and $\frac{\partial }{\partial \pi^{*i}} = \frac{1}{2}
(\frac{\partial }{\partial \Re{\pi^i}} + i \frac{\partial }{\partial
\Im{\pi^i}})$. In other words it is the tangent vector to $CP(N-1)$
\begin{eqnarray}
H=T_c+T_q +V_b = \hbar \Omega^c \Phi_c^i \frac{\partial }{\partial
\pi^i} + \hbar \Omega^q \Phi_q^i \frac{\partial }{\partial \pi^i} +
\hbar \Omega^b \Phi_b^i \frac{\partial }{\partial \pi^i} + c.c.
\label{field}
\end{eqnarray}
The characteristic equations for the PDE $H|E>=E|E>$ give the
parametric representations of their solutions in $CP(N-1)$. We will
identify the parameter $\tau$ in these equations with a ``universal
time of evolution'' such as the world time \cite{H1}. This time is
the measure of the configuration variation, i.e. it is a measure of
the distance in $CP(N-1)$ (an evolution trajectory length in the
Fubini-Study metric) expressed in time units. The energy
quantization will be discussed elsewhere.

\section{Dynamical quantum space-time}
We shall now construct a dynamical notion of space-time in terms of
internal quantum amplitudes. In this way we shall arrive at an
intrinsic definition of state-dependent space-time, consistently
energies from the dynamics of a quantum system. In the limit of weak
interaction, with essentially free motion, one can find an
approximate correspondence with the classical idea of space-time.

There are two key ideas presented here:

1. A conservation law (parallel transport) of a local Hamiltonian as
characterization of particles (excitations of GCS);

2. Identification of ``Lorentz spin transformation matrix" \cite{G}
of the spinor and classical Lorentz transformations of an inertial
frame.

Let us introduce the concept of a ``dynamical space-time'' as a new
construction capable of detecting the coincidences of the spinor
components in the formal two-level ``detector'' which is a part of
the full quantum configuration. The realization of this ``detector''
is of course the free choice of an observer. It is important only
that the chosen LDV should be invariantly connected with the
coherent state with respect to one of the points of the LDV spectra.
Let us assume this spectra of the LDV are known.

\subsection{Embedding ``Hilbert (quantum)
dynamics" in space-time} If we would like to have some embedding of
the ``Hilbert (quantum) dynamics" in space-time we should to
formalize the quantum observation (or measurement of some internal
dynamical variable). This diffeomorphism between rays of $CP(N-1)$
and $SU(N)$ generators mentioned above, are realized in terms of the
local $SL(2,C)$ action onto the states space $C^2$ as follows. The
basis of these spaces form two types of vectors: the normal vector
$|N>$ to the tangent space at some point of $CP(N-1)$ corresponding
to eigenvalue $\lambda_D$ of a measuring dynamical variable
$\hat{D}$ and the tangent vector $|T>$, generated by the coset
generators of $G/H$. These describe the interaction used for the
measurement process. It is important to understand that the
measurement i.e. comparison of the expected spinor
$(\alpha_0,\beta_0)$ and the measured spinor $(\alpha_1,\beta_1)$
pave the way to embedding Hilbert space dynamics into the local
dynamical space-time. One has a two-level system (logical spin $1/2$
\cite{Le4}) created by the quantum question (non-self-adjoint)
projector onto one of the two states $|N>,|T>$. Their coherent
states are given by the spinors $(\alpha,\beta)$ connected with
infinitesimal $SL(2,C)$ transformations, giving rise to the
variation of the space-time coordinates generated by local
infinitesimal Lorentz transformations.

The LDV is a vector field defined over $CP(N-1)$ and the comparison
of the LDV's at different GCS's (initial and perturbed due to
interaction used for measurement) require some procedure for the
{\it identification}. It is impossible to compare expected and
measured LDV ``directly'' (lack of correspondence due to curvature
of the $CP(N-1)$ \cite{Le3}). Affine parallel transport is
sufficient for this purpose. Parallel transport forms the condition
for the coefficient functions of the LDV leading to nonlinear field
equations in local dynamical space-time.

\subsection{Differential geometry of the measuring procedure}
The numerical value of some observable depends as a rule on a setup
(the character of  motion of laboratory, type of the measuring
device, field strength, etc.). However the relationships between
numerical values of dynamical variables and numerical
characteristics of laboratory motion, field strength, etc., should
be formulated in an invariant way, since they reflect the objective
character of the physical interaction used in the measurement
process. The numbers obtained  due to the measurements carry
information which does not exist a priori, i.e. before the
measurement process.

POSTULATE 3

{\it The invariant i.e. physically essential part of information
represented by the coherent states of the "logical spin 1/2" is
related to the space-time structure.} Such a postulate is based on
the observation that locally space-time is the Lorentz-invariant
manifold of points modeling different physical systems corresponding
to events depleted of all physical characteristics. In principle
arbitrary local coordinates may be attributed to these points. The
spinor structure of the Lorentz transformations represents the
transformations of the coherent states of the "logical spin 1/2".
Thereby we can assume the measurement of the quantum dynamical
variables expressed by the spinor ``creates'' the local space-time
coordinates. We will formulate non-linear field equations in this
local space-time with the help of a variational principle referring
to the generator of the quantum state deformation.

The internal dynamics of the quantum configuration given by the
action amplitude should be somehow reflected in physical space-time.
We shall solve the ``inverse representation problem'': to find a
locally unitary representation of the dynamical group $SU(N)$ in the
dynamical space-time where the induced realization of the coherence
group $SU(2)$ of the spinor acts \cite{Le1,Le2}. Its components are
subject to the ``Lorentz spin matrix transformations'' \cite{G}. We
then build the local spinor basis invariantly related to the
generalized coherent state in the $CP(N-1)$ manifold. First of all
we have to have the local reference frame as some type of
``representation'' of $SU(N)$. Each local reference frame and,
hence, $SU(N)$ ``representation'', may be marked by the local
coordinates (\ref{coor}). Now we should almost literally repeat
differential geometry of a smooth manifold embedded in flat ambient
Hilbert space${\cal{H}}=C^N$. The geometry of this smooth manifold
is the projective Hilbert space equipped with the Fubini-Study
metric (\ref{FS}) and with the affine connection (\ref{Gamma}).

The contact between this abstract formulation of the quantum
dynamics and ordinary (linear) expressions for some dynamical
variable $\hat{D}$ in ${\cal{H}}=C^\infty=l_2$ may be established
due to following formulas:
\begin{equation}
\hat{D}=\sum_{a,b \geq 0} <a|\hat{D}|b> \hat{P}_{ab}=\sum_{a,b \geq
0}D_{ab}\hat{P}_{ab}=\sum_{a,b \geq 0} F^{\alpha}_D
\hat{\lambda}_{\alpha,(ab)} \hat{P}_{ab},
\end{equation}
where projector
\begin{equation}
\hat{P}_{ab}=|a><b|=\frac{1}{\sqrt{a! b!}}:(\eta^+)^a \exp(-\eta^+
\eta)(\eta)^a:,
\end{equation}
where $:...:$ means the the normal ordering of operators, and
functions $F^{\alpha}_D$ obey some field equations which will be
discussed later. In particular, the Hamiltonian has a similar
representation
\begin{equation}
\hat{H}=\sum_{a,b \geq 0} <a|\hat{H}|b> \hat{P}_{ab}=\sum_{a,b \geq
0}H_{ab}\hat{P}_{ab}=\hbar \sum_{a,b \geq 0} \Omega^{\alpha}
\hat{\lambda}_{\alpha,(ab)} \hat{P}_{ab},
\end{equation}
i.e $F^{\alpha}_H=\hbar \Omega^{\alpha}$ \cite{Le8}. In order to
express the measurement of the ``particle's field'' in geometrically
intrinsic terms, we assume that the GCS is expressed in local
coordinates is
\begin{equation}
|G(\pi^1,...,\pi^{N-1})>=\sum_0^{N-1} g^a(\pi)|a>,
\end{equation}
where
\begin{eqnarray}
g^0(\pi^1,...,\pi^{N-1})=\frac{R^2}{\sqrt{R^2+\sum_{s=1}^{N-1}|\pi^s|^2}},
\end{eqnarray}
and for $1\leq i\leq N-1$ one has
\begin{eqnarray}
g^i(\pi^1,...,\pi^{N-1})=\frac{R
\pi^i}{\sqrt{R^2+\sum_{s=1}^{N-1}|\pi^s|^2}},
\end{eqnarray}
i.e. $CP(N-1)$ is embedded in the Hilbert space ${\cal{H}}=C^N$.

Then the rate of the ground state evolution in world time is the
tangent vector $|H>$ to the evolution curve $\pi^i=\pi^i(\tau)$
given by the formula
\begin{eqnarray}\label{41}
|H> = \frac{d|G>}{d\tau}=\frac{\partial g^a}{\partial
\pi^i}\frac{d\pi^i}{d\tau}|a\hbar>=|T_i>\frac{d\pi^i}{d\tau}=H^i|T_i>,
\end{eqnarray}
where
\begin{eqnarray}\label{42}
|T_i> = \frac{\partial g^a}{\partial \pi^i}|a\hbar>=T^a_i|a\hbar>.
\end{eqnarray}
Then the ``acceleration'' is
\begin{eqnarray}\label{43}
|A> =
\frac{d^2|G>}{d\tau^2}=|g_{ik}>\frac{d\pi^i}{d\tau}\frac{d\pi^k}{d\tau}
+|T_i>\frac{d^2\pi^i}{d\tau^2}=|N_{ik}>\frac{d\pi^i}{d\tau}\frac{d\pi^k}{d\tau}\cr
+(\frac{d^2\pi^s}{d\tau^2}+\Gamma_{ik}^s
\frac{d\pi^i}{d\tau}\frac{d\pi^k}{d\tau})|T_s>,
\end{eqnarray}
where
\begin{eqnarray}\label{44}
|g_{ik}>=\frac{\partial^2 g^a}{\partial \pi^i \partial \pi^k}
|a\hbar>=|N_{ik}>+\Gamma_{ik}^s|T_s>
\end{eqnarray}
and state
\begin{eqnarray}\label{45}
|N> = N^a|a\hbar>=(\frac{\partial^2 g^a}{\partial \pi^i \partial
\pi^k}-\Gamma_{ik}^s \frac{\partial g^a}{\partial \pi^s})
\frac{d\pi^i}{d\tau}\frac{d\pi^k}{d\tau}|a\hbar>
\end{eqnarray}
is normal to the ``hypersurface'' of the ground states. Then the
minimization of this ``acceleration'' under the transition from
point $\tau$ to $\tau+d\tau$ may be achieved by the annihilation of
the tangential component
\begin{equation}
(\frac{d^2\pi^s}{d\tau^2}+\Gamma_{ik}^s
\frac{d\pi^i}{d\tau}\frac{d\pi^k}{d\tau})|T_s>=0
\end{equation}
i.e. under the condition of affine parallel transport of the
Hamiltonian vector field
\begin{equation}\label{par_tr}
dH^s +\Gamma^s_{ik}H^id\pi^k =0.
\end{equation}

We use the Gauss-Codazzi equations
\begin{eqnarray}\label{46}
\frac{\partial N^a}{\partial \pi^i}=B^s_i T^a_s \cr \frac{\partial
T_k^a}{\partial \pi^i}-\Gamma^s_{ik}T^a_s=B_{ik}N^a
\end{eqnarray}
as a condition of integrability. These give us the dynamics of the
vacuum (normal) vector and the tangent vectors, i.e. one has the
local reference frame dynamics modeling the ``moving
representation''
\begin{eqnarray}\label{47}
\frac{d N^a}{d \tau}=\frac{\partial N^a}{\partial \pi^i} \frac{d
\pi^i}{d\tau}+c.c.= B^s_i T^a_s \frac{d \pi^i}{d\tau} +c.c. = B^s_i
T^a_s H^i +c.c.; \cr \frac{d T_k^a}{d \tau}=\frac{\partial
T_k^a}{\partial \pi^i}\frac{d \pi^i}{d\tau} +c.c. =
(B_{ik}N^a+\Gamma^s_{ik}T^a_s)\frac{d \pi^i}{d\tau}+c.c. \cr =
(B_{ik}N^a+\Gamma^s_{ik}T^a_s) H^i+c.c.
\end{eqnarray}
Note, that $0 \leq a \leq N$, but $1\leq i,k,m,n,s \leq N-1$. The
tensor $B_{ik}$ of the second quadratic form of the ground states
``hypersurface'' is as follows:
\begin{eqnarray}\label{48}
B_{ik} =<N|\frac{\partial^2 }{\partial \pi^i \partial \pi^k}|G>.
\end{eqnarray}

Now one should build the spinor in the local basis $(|N>,|D>)$ for
the quantum question with respect to the measurement of some local
dynamical variable $\vec{D}$. We will assume that there is {\it
natural state $|\widetilde{D}>$ of the quantum system in the local
reference frame representation} equal to the lifting of LDV $\vec{D}
\in T_{\pi}CP(N-1)$ into the Hilbert space $\cal{H}$, and there is
{\it expected state}
$|D_{expect}>=\alpha_0|N>+\beta_0|\widetilde{D}>$, associated with
the tuning of a ``measuring device''. This measuring device is
associated with the local projector along the normal $|N>$ onto the
natural state $\widetilde{|D>}$. In fact it defines the covariant
derivative in $CP(N-1)$. The lift-vectors $|N>,|D>$ are given by the
solutions of (\ref{47}) arising under interaction used for the
measurement of the LDV $\vec{D}$. In general $|D>$ is not a tangent
vector to $CP(N-1)$. But the normalized vector defined as the
covariant derivative $|\widetilde{D}>=|D>-<Norm|D>|Norm>$ is a
tangent vector to $CP(N-1)$ (it is convenient to take
$|Norm>=\frac{|N>}{\sqrt{<N|N>}}$). The operation of the
$|\widetilde{D}>$ normalization is a projector. Indeed,
\begin{eqnarray}
\widetilde{|\widetilde{D}>}= \widetilde{|D>-<Norm|D>|Norm>}\cr =
|D>-<Norm|D>|Norm> \cr - <Norm|(|D>-<Norm|D>|Norm>)|Norm> \cr
=|D>-<Norm|D>|Norm> = |\widetilde{D}>.
\end{eqnarray}
Then at the point $(\pi^1,...,\pi^{N-1})$ one has two components of
the spinor
\begin{eqnarray}\label{513}
\alpha_{(\pi^1,...,\pi^{N-1})}=\frac{<N|D_{expect}>}{<N|N>} \cr
\beta_{(\pi^1,...,\pi^{N-1})}=\frac{<\widetilde{D}|D_{expect}>}
{<\widetilde{D}|\widetilde{D}>}
\end{eqnarray}
then at an infinitesimally close point
$(\pi^1+\delta^1,...,\pi^{N-1}+\delta^{N-1})$ one has the new spinor
\begin{eqnarray}\label{514}
\alpha_{(\pi^1+\delta^1,...,\pi^{N-1}+\delta^{N-1})}=\frac{<N'|D_{expect}>}
{<N'|N'>} \cr \beta_{(\pi^1+\delta^1,...,\pi^{N-1}+\delta^{N-1})}=
\frac{<\widetilde{D}'|D_{expect}>}{<\widetilde{D}'|\widetilde{D}'>}
\end{eqnarray}
where the basis $(|N'>,|\widetilde{D}'>)$ is the lift of the
parallel transported $(|N>,|\widetilde{D}>)$ from the
infinitesimally close $(\pi^1+\delta^1,...,\pi^{N-1}+\delta^{N-1})$
back to $(\pi^1,...,\pi^{N-1})$.

These two infinitesimally close spinors being expressed as functions
of $\theta,\phi,\psi,R$ and
$\theta+\epsilon_1,\phi+\epsilon_2,\psi+\epsilon_3,R+\epsilon_4,$
may be represented as follows
\begin{eqnarray}\label{s1}
\eta = R \left( \begin {array}{c} \cos \frac{\theta}{2}(\cos
\frac{\phi_1- \psi_1}{2} - i\sin \frac{\phi_1 - \psi_1}{2}) \cr \sin
\frac{\theta}{2} (\cos \frac{\phi_1+\psi_1}{2} +i \sin
 \frac{\phi_1+\psi_1}{2})  \end {array}
 \right)
 = R\left( \begin {array}{c} C(c-is) \cr S( c_1+is_1)
\end
{array} \right)
\end{eqnarray}
and
\begin{eqnarray}
& \eta+\delta \eta = R\left( \begin {array}{c} C(c-is) \cr S(
c_1+is_1) \end {array} \right) \cr &+ R\left( \begin {array}{c}
S(is-c)\epsilon_1-C(s+i c)\epsilon_2+
C(s+ic)\epsilon_3+C(c-is)\frac{\epsilon_4}{R} \cr
 C(c_1+is_1)\epsilon_1+S(ic_1-s_1)\epsilon_2-S(s_1-ic_1)\epsilon_3
+S(c_1+is_1)\frac{\epsilon_4}{R}
\end
{array} \right).
\end{eqnarray}
They may be connected with an infinitesimal ``Lorentz spin
transformations matrix'' \cite{G}
\begin{eqnarray}
L=\left( \begin {array}{cc} 1-\frac{i}{2}\tau ( \omega_3+ia_3 )
&-\frac{i}{2}\tau ( \omega_1+ia_1 -i ( \omega_2+ia_2)) \cr
-\frac{i}{2}\tau
 ( \omega_1+ia_1+i ( \omega_2+ia_2))
 &1-\frac{i}{2}\tau( -\omega_3-ia_3)
\end {array} \right).
\end{eqnarray}
Then accelerations $a_1,a_2,a_3$ and angle velocities $\omega_1,
\omega_2, \omega_3$ may be found in the linear approximation from
the equation
\begin{equation}\label{equ}
\eta + \delta \eta = L \eta
\end{equation}
as functions of the spinor components depending on local coordinates
$(\pi^1,...,\pi^{N-1})$. Hence infinitesimal Lorentz transformations
define small ``space-time'' shifts. It is convenient to take Lorentz
transformations in the following form $ct'=ct+(\vec{x} \vec{a})
d\tau, \quad \vec{x'}=\vec{x}+ct\vec{a} d\tau +(\vec{\omega} \times
\vec{x}) d\tau$, where we put $\vec{a}=(a_1/c,a_2/c,a_3/c), \quad
\vec{\omega}=(\omega_1,\omega_2,\omega_3)$ \cite{G} in order to have
for $\tau$ the physical dimension of time. The coordinates $x^\mu$
of points in this space-time serve in fact for the parametrization
of deformations of a ``field shell'' arising under its motion
according to non-linear field equations \cite{Le1,Le2}.

\section{Field equations in the dynamical space-time}
In order to find the ``field shell'' of the perturbed GCS one should
establish some wave equations in the dynamical space-time. All these
notions require more precise definitions. Namely, say, in the
simplest case of $CP(1)$, the ``field shell'' represented in the
spherical coordinates is the classical vector field
$\Omega^{\alpha}=\frac{x^{\alpha}}{r}(\omega +i \gamma), \quad 1\leq
\alpha \leq 3 $ giving the rate of the GCS variations. The tensor
fields $1\leq \alpha \leq 8,15,...,N^2-1$ will be discussed
elsewhere. Note, that quanta numbers are now strongly connected with
the tensor character of the GCS driving field $\Omega^{\alpha}$.
These fields are ``classical'' since they are not subject to
quantization directly, i.e. by the attribution of the fermionic or
bosonic commutation relations. They obey nonlinear field equations
having soliton-like solutions. Their internal dynamical variables
like spin, charge, etc.,  are a consequence of their dynamical
structure.

A ``particle'' which may be associated with the ``field shell'' in
the dynamical space-time (see below), is now described locally by
the Hamiltonian vector field $H$. At each point
$(\pi^1,...,\pi^{N-1})$ of the $CP(N-1)$ one has an ``expectation
value'' of the $H$ defined by a measuring device. But a displaced
GCS may by reached along one of the continuum paths. Therefore the
comparison of two vector fields and their ``expectation values'' at
neighboring points requires some natural rule. The comparison makes
sense only for the same ``particle'' or for its ``field shell''
along some path. For this reason one should have an identification
procedure. The affine parallel transport in $CP(N-1)$ of vector
fields is a natural and the simplest rule for the comparison of
corresponding ``field shells''. Physically the identification of
``particle'' literally means that its Hamiltonian vector field is a
Fubini-Study covariant constant.

Since we have only the unitary fields $\Omega^{\alpha}$ as
parameters of the GCS transformations we assume that in accordance
with an equivalence-like principle under the infinitesimal shift of
the unitary field $\delta \Omega^{\alpha}$ in the dynamical
space-time, the shifted Hamiltonian field should coincide with the
infinitesimal shift of the tangent Hamiltonian field generated by
the parallel transport in $CP(N-1)$  during world time $\delta \tau$
\cite{H1}. Thus one has
\begin{equation}
\hbar (\Omega^{\alpha} + \delta \Omega^{\alpha} ) \Phi^k_{\alpha}=
\hbar \Omega^{\alpha}( \Phi^k_{\alpha} - \Gamma^k_{mn}
\Phi^m_{\alpha} V^n \delta \tau)
\end{equation}
and, hence, in accordance with the sufficiency criterion for the
equation with non-trivial solution
\begin{equation}
 \hbar(\delta \Omega^{\alpha} \delta^k_m+
 \Omega^{\alpha} \Gamma^k_{mn} V^n \delta \tau) \Phi^m_{\alpha}=0
\end{equation}
one has following equations
\begin{equation}
\frac{ \delta \Omega^{\alpha}}{\delta \tau} = -
\Omega^{\alpha}\Gamma^m_{mn} V^n,
\end{equation}
(there is no  summation in $m$). We introduce the dynamical
space-time coordinates $x^{\mu}$ as state-dependent quantities,
transforming in accordance with the local Lorentz transformations
$x^{\mu} + \delta x^{\mu} = (\delta^{\mu}_{\nu} +
\Lambda^{\mu}_{\nu} \delta \tau )x^{\nu}$. The parameters of
$\Lambda^{\mu}_{\nu} (\pi^1,...,\pi^{N-1})$ depend on the local
transformations of local reference frame in $CP(N-1)$ described in

the previous paragraph. Assuming a spherically symmetrical solution,
we will use the coordinates $(x^0=ct,x^1=r\sin \Theta \cos \Phi,
x^2=r\sin \Theta \sin \Phi, x^3=r\cos \Theta)$. In the case of
spherical symmetry, $\Omega^1=(\omega+i \gamma) \sin \Theta \cos
\Phi, \Omega^2=(\omega+i \gamma) \sin \Theta \sin \Phi,
\Omega^3=(\omega+i \gamma) \cos \Theta)$ and in the general case of
the separability of the angle and radial parts, one has
$\Omega^{\alpha}=\sum C_{l,m}^{\alpha} Y_{l,m}(\Theta,\Phi)(\omega+i
\gamma)=\sum C_{l,m}^{\alpha} Y_{l,m}(\Theta,\Phi)\Omega$. Then
taking into account the expressions for the ``4-velocity"
$v^{\mu}=\frac{\delta x^{\mu}}{\delta \tau} = \Lambda^{\mu}_{\nu}
(\pi^1,...,\pi^{N-1}) x^{\nu} $ one has the field equation
\begin{equation}\label{FSE}
v^{\mu} \frac{\partial \Omega}{\partial x^{\mu} } = -
\Omega\Gamma^m_{mn} V^n,
\end{equation}
where
\begin{equation}
\matrix{ v^0&=&(\vec{x} \vec{a}) \cr
 \vec{v}&=&ct\vec{a}  +(\vec{\omega} \times
\vec{x}) \cr }.
\end{equation}
or, in detail,
\begin{equation}
\matrix{ v^0&=&r(a_1 \sin \Theta \cos \Phi+a_2\sin \Theta \sin \Phi+
a_3\cos \Theta ) \cr
 v^1&=&cta_1 +r(\omega_2 \cos
\Theta-\omega_3\sin \Theta \sin \Phi) \cr v^2&=&cta_2 +r(\omega_3
\sin \Theta \cos \Phi-\omega_1 \cos \Theta ) \cr v^3&=&cta_3
+r(\omega_1 \sin \Theta \sin \Phi-\omega_2\sin \Theta \cos \Phi) \cr
}.
\end{equation}

If one wishes to find a field corresponding to a given trajectory,
say, a geodesic in $CP(N-1)$, then, taking into account that any
geodesic as whole belongs to some $CP(1)$, one may put $ \pi^1=
e^{i\phi} \tan(\sigma \tau)$. Then $V^1=\frac{d \pi^1}{d
\tau}=\sigma \sec^2(\sigma \tau) e^{i\phi}$, and one has a linear
wave equations for the gauge unitary field $\Omega^{\alpha}$ in the
dynamical space-time with complicated coefficient functions of the
local coordinates $(\pi^1,...,\pi^{N-1})$.  Under the assumption
$\tau = w t$ this equation has following solution
\begin{eqnarray}
& \omega+i \gamma \cr &=(F_1(r^2-c^2t^2)+i F_2(r^2-c^2t^2))
\exp{(-2w c \int_0^t dp \frac{ \tan(w p)}{A
\sqrt{c^2(p^2-t^2)+r^2}})},
\end{eqnarray}
where $F_1,F_2$ are an arbitrary functions of the interval
$s^2=r^2-c^2t^2$, $(\vec{a},\vec{x})=A r \cos(\chi)$,
$A=\sqrt{a_x^2+a_y^2+a_z^2}$ and $r=\sqrt{x^2+y^2+z^2}$. The angle
$\chi$ in fact is defined by a solution of the equation (4.20). We
used $\chi=\pi$ since for us it is now interesting only (due to
spherical symmetry) a ``radial boost toward the center of the field
shell".

The general factor demonstrates the spreading of the light cone due
to the boosts. Thus our results are consistent with the so-called
``off-shell'' idea of Horwitz-Piron-Stueckelberg \cite{HP,H2}.
Namely, state-dependent Lorentz transformations intended to reach
the coincidence of two infinitesimally close spinors lead to
absolute acceleration (in the sense of deformation of quantum
state). This is a measure of the deformation of GCS's \cite{Le6}.
One may treat this as an effective self-interaction field of GCS
leading to the spreading of the mass shell. This lump might serve as
non-singular source of the field in QFT.

It is interesting to analyze the behavior of the general factor in
the vicinity of the light-cone and in the remote areas. For this
purpose it is convenient to put $w=1,A=1,c=1$ and, using new
variable $q=p/t $ one has
\begin{equation}
I(t,r)=\int_0^t dp \frac{ \tan(p)}{ \sqrt{p^2-t^2+r^2}})=\int_0^1 dq
\frac{ \tan(tq)}{\sqrt{\frac{r^2}{t^2}-1+q^2}}.
\end{equation}
Let us put $a^2=\frac{r^2}{t^2}-1$. In the vicinity of the
light-cone $a^2 \ll 1$ the vicinity of $q=0$ gives the main
contribution to the integral and, hence,  $\tan(tq) \simeq tq$,
therefore
\begin{equation}
I(t,r) \simeq I_0(t,r)= \int_0^1 dq \frac{
tq}{\sqrt{a^2+q^2}}=r-\sqrt{r^2-t^2}.
\end{equation}
The behavior of the factor $\exp(-2(r-\sqrt{r^2-t^2})) $ is depicted
in the Fig.1. \vskip .1cm
\begin{figure}
\includegraphics[width=4in]{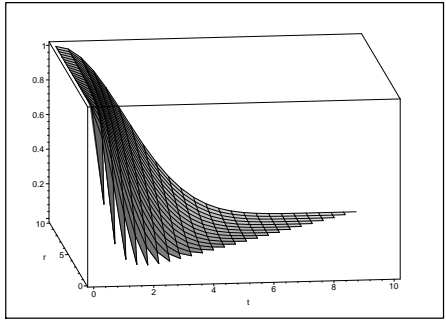}
\caption{The factor $\exp(-2(r-\sqrt{r^2-t^2})) $ in the vicinity of
the light-cone } \label{fig.1}
\end{figure}
\vskip .1cm On the other hand, if $a^2 \gg 1$, then
\begin{equation}
I(t,r) \simeq I_r(t,r)= \frac{1}{a}\int_0^1 \tan(tq) dq =
\frac{\tan(t)^2}{\sqrt{r^2-t^2}},
\end{equation}
and the behavior of the factor $ \exp(-2
\frac{\tan(t)^2}{\sqrt{r^2-t^2}}) $ is depicted in the Fig.2. It is
seen that for small $t,r$ the behavior of these factors is similar.
\vskip .1cm
\begin{figure}
\includegraphics[width=4in]{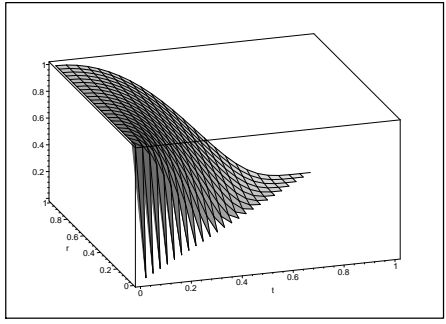}
\caption{The factor $\exp(-2 \frac{\tan(t)^2}{\sqrt{r^2-t^2}})$ in
the remote areas from the light-cone } \label{fig.2}
\end{figure}
\vskip .1cm

Let us discuss now the self-consistent problem
\begin{equation}
v^{\mu} \frac{\partial \Omega^{\alpha}}{\partial x^{\mu} } = -
(\Gamma^m_{mn} \Phi_{\beta}^n+\frac{\partial
\Phi_{\beta}^n}{\partial \pi^n}) \Omega^{\alpha}\Omega^{\beta},
\quad \frac{d\pi^k}{d\tau}= \Phi_{\beta}^k \Omega^{\beta}
\end{equation}
arising under the condition of parallel transport
\begin{eqnarray}
\frac{\delta H^k}{\delta \tau} &= &\hbar \frac{\delta
(\Phi^k_{\alpha} \Omega^{\alpha})}{\delta \tau}=0
\end{eqnarray}
of the Hamiltonian field (\ref{par_tr}), \cite{Le7,Le8}. Again we
will assume the simplest case of $CP(1)$ dynamics when $1\leq
\alpha,\beta \leq3,\quad i,k,n=1$. This system being split into the
real and imaginary parts takes the form
\begin{eqnarray}
\matrix{ (r/c)\omega_t+ct\omega_r=-2\omega \gamma F(u,v), \cr
(r/c)\gamma_t+ct\gamma_r=(\omega^2 - \gamma^2) F(u,v), \cr u_t=\beta
U(u,v,\omega,\gamma), \cr v_t=\beta V(u,v,\omega,\gamma), }
\label{self_sys}
\end{eqnarray}
where
\begin{eqnarray}
\matrix{ \pi(t,r)=u(t,r)+iv(t,r), \cr F(u,v)=(2\sin\Theta\cos\Phi
u(t,r)+2\sin\Theta\sin\Phi v(t,r)+ \cr
\cos\Theta(1-u(t,r)^2-v(t,r)^2)\frac{r \cos
\Theta}{(1+u(t,r)^2+v(t,r)^2)(\vec{x}\vec{a})}, \cr
U(u,v,\omega,\gamma)=(1/2)\sin\Theta
\cos\Phi\gamma(-1+u(t,r)^2-v(t,r)^2)+\cr \omega(\sin\Theta\cos\Phi
v(t,r)u(t,r)-(1/2)\sin\Theta\sin\Phi(1-u(t,r)^2+v(t,r)^2))+\cr
\gamma(\sin\Theta\sin\Phi v(t,r)u(t,r)+\cos\Theta
u(t,r))+\omega\cos\Theta v(t,r),\cr V(u,v,\omega,\gamma)=
(1/2)\sin\Theta \cos\Phi\omega(1-u(t,r)^2+v(t,r)^2)+\cr
\gamma(\sin\Theta\cos\Phi
v(t,r)u(t,r)-(1/2)\sin\Theta\sin\Phi(1+u(t,r)^2-v(t,r)^2))-\cr
\omega(\sin\Theta\sin\Phi v(t,r)u(t,r)+\cos\Theta
u(t,r))+\gamma\cos\Theta v(t,r) }.
\end{eqnarray}
It is impossible of course to solve this self-consistent problem
analytically even in this simplest case of the two state system, but
it is reasonable to develop a numerical approximation in the
vicinity of the following exact solution. Let us put $\omega=\rho
\cos \psi, \quad \gamma=\rho \sin \psi$, then, assuming for
simplicity that $\omega^2+\gamma^2=\rho^2=constant$, the two first
PDE's may be rewritten as follows:
\begin{equation}
\frac{r}{c}\psi_t+ct\psi_r=F(u,v) \rho \cos \psi.
\end{equation}
The exact solution of this quasi-linear PDE is
\begin{equation}
\psi_{exact}(t,r)=\arctan \frac{\exp(2c\rho F(u,v)
f(r^2-c^2t^2))(ct+r)^{2F(u,v)}-1}{\exp(2c\rho F(u,v)
f(r^2-c^2t^2))(ct+r)^{2F(u,v)}+1}, \label{ex_sol}
\end{equation}
where $f(r^2-c^2t^2)$ is an arbitrary function of the interval. It
is useful to see the sketch of the spatial behavior of this
solution. We have put $c=\rho=F(u,v)=1, \quad  f(r^2-c^2t^2)=-
(r^2-c^2t^2)^2$. The 3D graphics of amplitudes
$\omega(t,r),\gamma(t,r)$ and corresponding spatial fields are
depicted in Fig 3-6. The real part is an isotropic vector field;
somewhat analogous to a charge, whereas the imaginary part is an
isotropic but that looks like a bubble in the force field. The
physical interpretation of this solution is still an open question.
Nevertheless, it is possible to give some general interpretation
which, of course, requires careful investigation. It is known that
quantum measurement induces some topologically non-trivial
monopole-like vector potential associated with a Berry geometric
phase \cite{A2}. The question is: is it possible to find a realistic
fundamental gauge potential if one uses not artificial parameter
space, but an inherently related projective Hilbert space $CP(N-1)$.
If we treat the acceleration parameter $A=1$ as an interaction
constant, then by (45) one may interpret $S(t,r)$ as some
relativistic source of a force field. \vskip .1cm
\begin{figure}
\includegraphics[width=4in]{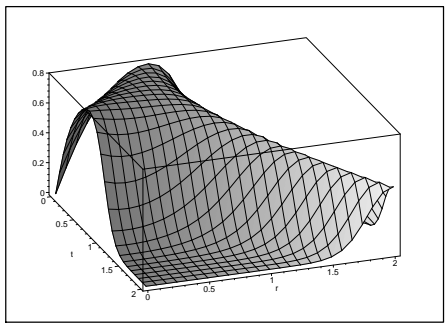}
\caption{The real part $\omega(t,r)$ of the amplitude} \label{fig.3}
\end{figure}
\vskip .2cm
\begin{figure}
\includegraphics[width=4in]{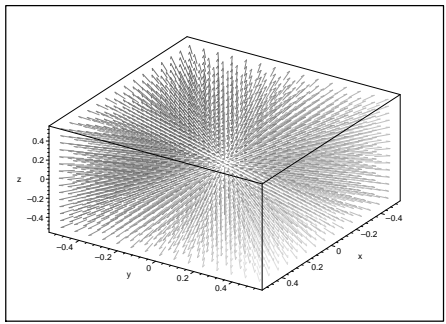} \caption{The
real part  $\frac{x^\alpha}{r}\omega(t=1,r)$ of the spatial field  }
\label{fig.4}
\end{figure}
\vskip .2cm
\begin{figure}
\includegraphics[width=4in]{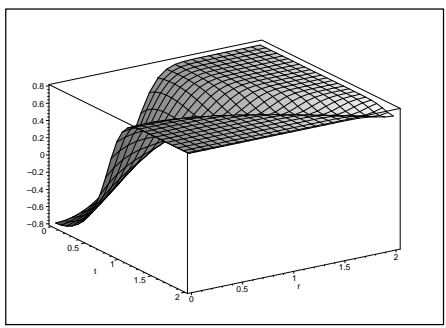} \caption{The
imaginary part $\gamma(t,r)$ of the amplitude} \label{fig.5}
\end{figure}
\vskip .1cm
\begin{figure}
\includegraphics[width=4in]{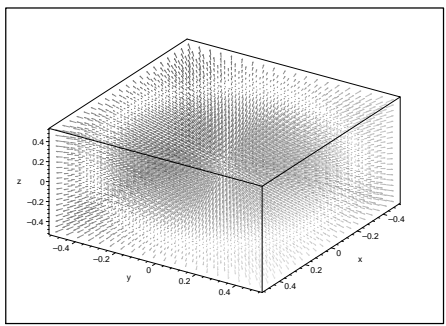} \caption{The
imaginary  part $\frac{x^\alpha}{r}\gamma(t=1,r)$ of the spatial
field} \label{fig.6}
\end{figure}
\vskip .1cm
\section{Discussion}
It is clear that the {\it coincidence} of the pointer with some
number on the scale is in fact the coincidence of the two space-time
points. But one has in the quantum area literally neither pointer
nor the scale; instead we deal with a `` field shell". In Einstein's
original thinking \cite{Einstein1}, he had intuitively clear
classical measuring devices (clocks, scales, rods, etc.) and the
intuitively clear space-time coincidence of two ``points''. Without
these ingredients it is difficult to image classical measurement
processes and even the notion of space-time itself. However, in fact
{\it space-time separation is state-dependent} \cite{Einstein2,A1}.
In such a situation, one should decide on what criterion of identity
is physically acceptable. It has been formulated here as the affine
parallel transport of the local Hamiltonian vector field over
$CP(N-1)$.

A review of the proposed scheme is as follows:

a). We discuss the representation of the $G=SU(N)$ acting on the
states $|S>  \in {\cal{H}}$ in terms of local dynamical variables
representation by the tangent vectors to $CP(N-1)$ (the operators of
differentiation); these correspond to states of the action.

b). Measurement is realized as a perturbation of the generalized
coherent quantum state.

c). Identification of quantum systems with help of the affine
parallel transport agrees with Fubini-Study metric.

d). A variational principle applied to the local Hamiltonian leads
to quasi-linear PDE field equations for the ``field shell".

e). The fundamental ``field shell'' of the local dynamical variables
provides as a model of an extended particle serving for the
establishment of a local state-dependent space-time structure.

In the framework of our model the projection acts continuously and
locally along the $CP(N-1)$ trajectory of GCS onto the corresponding
tangent space, since it is the covariant differentiation of vector
fields representing LDV's on $CP(N-1)$.

Deformation of GCS occurs during interaction involved with
measurement. Let us discuss the dynamics of Schr\"odinger's lump
during measurement (see paragraph 30.10, \cite{Penrose}). This
construction is a humane version of the Schr\"odinger's cat. In
distinction with so complicated a system as a poisoned cat, and
indefinite displaced lump of matter, we would like discuss the
deformation of GCS which is theoretically analyzable.

First of all we should note that the assumption that ``the energy in
each case is the same" may be correct only approximately, say, in
the case of adiabatic ``kicking" of the lump. The finite time of
transition unavoidably leads to the acceleration of the lump of
matter, to the deformation of its quantum state \cite{Le6}, and to
the shift of mass-energy.

In the framework of our model, the GCS of the lump is ``kicked'' by
the coset transformations of $SU(N)$ group. The coefficient
functions of the $SU(N)$ generators obey some quasi-linear
relativistic field equations in the local dynamical space-time
\cite{Le1,Le2}.

The difference of the masses of the original and the displaced lumps
leads to different time-like Killing vectors (if any) in the
vicinities of two lumps. This is an obstacle to write Schr\"odinger
equations for superposed wave functions.

In the framework of our model we have state-dependent space-times
arising as specific sections of the tangent fibre bundle over
$CP(N-1)$. 

\end{document}